\documentclass[11pt]{article}
\usepackage{amsfonts,amssymb,amsmath,amsthm}

\usepackage{graphicx,color}

\newcommand{\ket}[1]{\left | #1 \right \rangle}
\newcommand{\bra}[1]{\left \langle #1 \right |}

\newcommand*{\cE}{\mathcal{E}}
\newcommand*{\cL}{\mathcal{L}}
\newcommand*{\complex}{\mathbb{C}}

\def\openone{\leavevmode\hbox{\small1\kern-3.8pt\normalsize1}}

\def\ce{{\cal E}}

\def\cm{{\cal M}}

\def\RR{\mathbb{R}}

\newtheorem{theorem}{Theorem}
\newtheorem{lemma}{Lemma}
\newtheorem{proposition}{Proposition}

\theoremstyle{definition}

\newtheorem{property}{Property}

\newcommand{\proj}[1]{\ket{#1}\!\bra{#1}}

\newcommand{\pder}[2]{\frac{\partial #1}{\partial #2 }}

\begin{document}
\title{\LARGE\bf
 Symmetric polynomials in information\\ theory: entropy and subentropy}
\author{Richard Jozsa
  and Graeme Mitchison\\[3mm]
  \small\it
  \small\it DAMTP, Centre for Mathematical Sciences, University of Cambridge,\\ \small\it Wilberforce Road, Cambridge CB3 0WA, U.K.}

\date{}

\maketitle

\begin{abstract} Entropy and other fundamental quantities of information theory are customarily expressed and manipulated as functions of probabilities. Here we study the entropy $H$ and subentropy $Q$ as functions of the elementary symmetric polynomials in the probabilities, and reveal a series of remarkable properties. Derivatives of all orders are shown to satisfy a complete monotonicity property. $H$ and $Q$ themselves become multivariate Bernstein functions and we derive the density functions of their Levy-Khintchine representations. We also show that $H$ and $Q$ are Pick functions in each symmetric polynomial variable separately. Furthermore we see that $H$ and the intrinsically quantum informational quantity $Q$ become surprisingly closely related in functional form, suggesting a special significance for the symmetric polynomials in quantum information theory. Using the symmetric polynomials we also derive a series of further properties of $H$ and $Q$.
\end{abstract}
\section{Introduction}\label{intro}
It is natural to represent the informational properties of
a random variable by functions that depend on the associated probabilities only as
an {\em unordered} set. Correspondingly, entropic functions that are
used as information measures are {\em symmetric} functions of the
probabilities. But we can incorporate this fundamental feature in a
deeper mathematical way: if we use the elementary symmetric
polynomials in the probabilities as primary variables, rather than the
probabilities themselves, then {\em arbitrary} functions will
automatically depend on the probabilities only as an unordered set. In
this paper we will study the Shannon entropy $H$ and subentropy $Q$
(cf below), as functions of the symmetric polynomials in the
probabilities. We will see that they both then exhibit a series of
remarkable properties. Furthermore the use of symmetric polynomials as
variables will reveal surprising relationships between $H$ and
the intrinsically quantum information theoretic quantity $Q$.

Let $\{ x_1,. \ldots , x_d\}$ be a probability distribution. The  associated elementary symmetric polynomials are defined by
\[ e_1=\sum_j x_j, \hspace{5mm} e_2=\sum_{i<j} x_ix_j, \hspace{5mm}
e_3=\sum_{i<j<k} x_ix_jx_k, \hspace{5mm}\ldots \] We will lift the
probability condition $e_1=\sum x_j =1$ and use $(e_1, e_2, \ldots ,
e_d)$ as independent variables. For non-negative $x_j$'s we obviously
have each $e_k\geq 0$ although not all $d$-tuples of non-negative
$e_k$ values arise in this way: given any $e_1, \ldots , e_d \geq 0$
the associated $x_j$'s are the roots of the polynomial equation
\begin{equation}\label{polypeq}
p(x)=x^d - e_1 x^{d-1} \ldots +(-1)^ke_kx^{d-k} \ldots +(-1)^d e_d = 0
\end{equation}
so they may be either real non-negative or complex (in which case they must occur in complex conjugate pairs), but they cannot be real and negative. Below, when $H$ and $Q$ (defined initially as functions of non-negative $x_j$'s) are expressed as functions of the $e_k$'s, they are analytic functions on the associated restricted region of $(e_1, \ldots , e_d)$-space and it will be natural and convenient to view them as functions on the full space of all non-negative real $e_k$'s by analytic continuation. Intriguingly, in terms of our original variables $x_i$, this amounts to extending the notion of probability to a certain complex domain.

The entropy $H$ and subentropy $Q$ are defined by the following symmetric functions of the $x_j$'s:
\begin{equation}\label{heq} H(x_1, \ldots ,x_d) =-\sum_{i=1}^d x_i \ln x_i \end{equation}
\begin{equation} \label{qdeq}
Q(x_1, \ldots ,x_d)= - \large\sum_{i=1}^d  \frac{x_i^d}{ \prod_{ j \ne i}(x_i - x_j)}\, \ln x_i .
\end{equation}
(For coincident $x_j$'s the subentropy is defined to be the corresponding limit as the $x_j$'s become equal, which is always finite.)
For $d$-dimensional quantum states $\rho$ we define $H(\rho)$ and $Q(\rho)$ to be the above functions applied to the eigenvalues of $\rho$.

The subentropy function was introduced in \cite{wkw0,jrw} where it was shown to have the following physical significance:\\
(a) For any quantum state $\rho$ let $\ce = \{q_i;\ket{\psi_i}\}$ be
an ensemble of pure states $\ket{\psi_i}$ with density matrix $\rho$
i.e. $q_i$ are probabilities and $\sum_i q_i \proj{\psi_i}=\rho$. The
accessible information $I_{\rm acc}(\ce)$ of $\ce$ is defined to be
the maximum amount of classical mutual information about $i$ that can
be obtained from any measurement on the pure states
$\ket{\psi_i}$. According to Holevo's theorem \cite{hol} the von
Neumann entropy $H(\rho)$ is an attainable upper bound on $I_{\rm
  acc}(\ce)$ as $\ce$ ranges over all pure state ensembles with
density matrix $\rho$. In \cite{jrw} it was shown that dually,
$Q(\rho)$ is an attainable {\em lower} bound (being attained for the
so-called Scrooge ensemble).\\
(b) Let $\cm$ be a complete von Neumann measurement in $d$ dimensions
with associated orthonormal basis $\{ \ket{e_i}\}$. For a quantum
state $\rho$ let $p_i=\bra{e_i}\rho\ket{e_i}$ be the measurement
probabilities and let $H_\cm =H(p_1,\ldots , p_d)$ be the Shannon
entropy of the output distribution. If $\langle H_\cm \rangle$ denotes
the Haar-uniform average over choices of measurement basis then
$Q(\rho )= \langle H_\cm \rangle -(\frac{1}{2}+\ldots +\frac{1}{d})$.

Although $H$ is a quantity that has fundamental significance in
classical information theory, the subentropy $Q$ appears to be an
intrinsically quantum construct with no known natural significance in
classical information theory. Nevertheless we will see that when expressed in terms of
the $e_k$'s as variables, $H$ and $Q$ become surprisingly closely
related   (cf  for example eqs. (\ref{qsum}) and (\ref{qder1})  below) suggesting a special
significance for the $e_k$ variables for quantum information theory.

Our use of elementary symmetric polynomials as variables appears to be entirely novel in information theory. The exploration here was motivated by our earlier work \cite{mj04} in which it was shown that $\partial H/\partial e_k >0$ for $k\geq 2$, as an intermediate step for developing an interpretation of Schumacher compression in terms of the geometry of Hilbert space. The monotonicity of these first derivatives was also established (by different means) in \cite{fannes} for other purposes. Below we will see that (amongst other properties), similar (alternating) monotonicity conditions in fact hold for all higher order derivatives of both $H$ and $Q$, and in particular the functions $\partial H/\partial e_k$ for $k\geq 2$, are completely monotone functions in all their variables. Furthermore we will see that on the space $
\{ (e_1, \ldots ,e_d): e_1=1 \mbox{ and } e_2, \ldots , e_d \geq 0 \}$, the functions $H$ and $Q$ are multi-variate Bernstein functions and single-variable Pick functions in each variable separately.

\section{Contour integrals and half-axis formulae}\label{contourint} 
To express $H$ and $Q$ as functions of the $e_k$'s we will use the implicit relation eq. (\ref{polypeq}) between the $x_j$'s and the $e_k$'s, together with Cauchy's integral formula of complex analysis. Writing
\begin{equation}\label{polyp}
p(z)=z^d - e_1 z^{d-1} \ldots +(-1)^ke_kz^{d-k} \ldots +(-1)^d e_d ,
\end{equation}
we note that $p'(z)/p(z)= \sum_{j=1}^d \frac{1}{z-x_j}$ where $x_1, \ldots ,x_d$ are the roots of $p$, and hence
\begin{equation}\label{hee}
  H(e_1, \ldots ,e_d)=-\frac{1}{2\pi i}\oint z \ln z \,\, \frac{p'(z)}{p(z)} \, dz .\end{equation}
Here the contour in the complex $z$-plane surrounds all the $x_j$ values but excludes the origin $z=0$. For the complex logarithm we always use the negative real axis as the branch cut and use the branch given by $\ln z = \ln |z|+i \arg z$ with $-\pi <\arg z <\pi$.

The case of subentropy is actually simpler: if $g(z)$ is any function that is holomorphic in and on the contour then Cauchy's residue theorem gives (for distinct $x_j$'s)
\begin{equation}\label{oint}
\sum_{j=1}^d \frac{g(x_j)}{\prod_{i\neq j} (x_j-x_i)}= \frac{1}{2\pi i} \oint \frac{g(z)}{(z-x_1)\ldots (z-x_d)} \, dz .\end{equation}
Setting $g(z)=-z^d\ln z$ and comparing with eq. (\ref{qdeq}) for $Q$ we immediately get
\begin{equation}\label{subhee}
Q=-\frac{1}{2\pi i} \oint \frac{z^{d}\ln z}{p(z)}\, dz.
\end{equation}

We now derive expressions for $H$ and $Q$ as real integrals on the
positive real axis. These will be obtained from the contour integrals
above by distorting the contour into a keyhole contour that excludes
the negative real axis, running above and below it at a small distance
$\epsilon$ between $z=-R\pm i \epsilon$ and $z=0\pm i\epsilon$,
looping around the origin $z=0$ in a circular arc of radius $\epsilon$
and being completed by a circular arc of large radius $R$, all
traversed counterclockwise. Then we will consider the limits $\epsilon
\rightarrow 0$ and $R\rightarrow \infty$.

Consider any integral of the form
\begin{equation}\label{genint}
\frac{1}{2\pi i}\oint h(z)\, \ln z \, dz \end{equation}
around the keyhole contour where $h(z)$ generally has poles (but not branch points) inside the contour. {\em Suppose} that the sections  of the contour integral on the two circular arcs each tend to zero as $\epsilon \rightarrow 0$ and $R\rightarrow \infty$ respectively. Then recalling that $\arg z$ tends to $+\pi$, resp. $-\pi$, as $z$ approaches the negative $z$ axis from above, resp. below, and $\ln z=\ln |z|+i\arg z$, we see that 
\begin{equation}\label{hint}
\frac{1}{2\pi i}\oint h(z)\, \ln z \, dz  = \int_0^\infty h(-\tau)\, d\tau \end{equation}
(where the negative axis has been labelled as $-\tau$ for $\tau>0$).

Now consider again our contour integral formula eq. (\ref{hee}) for entropy:
\[ H(e_1, \ldots ,e_d)=\frac{1}{2\pi i}\oint -z \ln z \,\, \frac{p'(z)}{p(z)} \, dz .\]
The integrand satisfies the desired vanishing condition for $\epsilon \rightarrow 0$ but not for $R\rightarrow \infty$ (the integrand diverging as $\ln R$ on that circular arc). To remedy this, consider modifying the integrand by adding two further terms:
\begin{align}\label{correcting-terms}
 \oint \left[ -z \ln z \,\, \frac{p'(z)}{p(z)}+A\ln z + B\frac{\ln z}{z-1}\,\right] \, dz 
\end{align}
which do not change the value of the integral (since for the two terms respectively, $\ln z$ is holomorphic throughout the whole region inside the contour, and its residue $\ln 1$ at $z=1$ is zero too). The integrand has the form $\frac{r_1(z)}{r_2(z)}\ln z$ where $r_1$ and $r_2$ are polynomials of degree $d+1$, and using eq. (\ref{polyp}) for $p(z)$ we see that if we set $A=d$ and $B=e_1$ then the two leading coefficients of $r_1$ become zero and the asymptotic condition on the circular arc $R\rightarrow \infty$ will be satisfied. Thus we have
\[ H= \frac{1}{2\pi i}\oint \ln z \left[  - \frac{zp'(z)}{p(z)} + \frac{e_1}{z-1}+d\, \right] \, dz \]
with the integral now satisfying both conditions on the circular arcs with $\epsilon \rightarrow 0$ and $R\rightarrow \infty$.

Finally we apply eq. (\ref{hint}) with $h(z)=(-zp'(z)/p(z)+e_1/(z-1)+d)$. Introducing 
\begin{align}
\label{qdef} q(\tau)=\tau^d+e_1\tau^{d-1}+\ldots +e_k\tau^{d-k}+\ldots  +e_d,
\end{align} 
which is the polynomial with roots $-x_1, -x_2, \ldots, -x_d$, we get the {\em half-axis formula}:
\begin{equation}\label{hhalf}
  H(e_1, \ldots ,e_d)=\int_0^\infty   \left[  -\frac{\tau q'(\tau)}{q(\tau)} - \frac{e_1}{\tau+1}+d\, \right] \, d\tau \end{equation}
(with the $e_k$ dependence appearing explicitly on substituting eq. (\ref{qdef}) for $q(\tau)$).
By writing the integrand as $\tau f'(\tau)$ and integrating by parts, we obtain the further formula
\begin{equation}\label{fansimple}
H(e_1, \ldots , e_d)=\int_0^\infty   \left[ \, \ln q(\tau) +(e_1-d)\ln \tau -e_1 \ln (\tau+1)\, \right] \, d\tau \end{equation}
which will be useful later in section \ref{picky}.
An alternative (more complicated) formula of this kind for $H$ in the case of $e_1=1$, was given in \cite{fannes}. 

For the case of subentropy we can apply the same techniques to the
contour integral formula eq. (\ref{subhee}) giving
\[ Q=\frac{1}{2\pi i} \oint  \ln z \left[ -\frac{z^d}{p(z)}+\frac{e_1}{z-1}+1\right] \,\, dz \]
with the integrand satisfying both circular arc conditions, and eq. (\ref{hint}) gives
\begin{equation} \label{qhalf} 
Q(e_1, \ldots ,e_d)=\int_0^\infty \left[ - \frac{\tau^d}{q(\tau)}-\frac{e_1}{(\tau+1)}+1 \right] \,\, d\tau.
\end{equation}

It is interesting to note a tantalising similarity between the formulae for $H$ and $Q$ in eqs. (\ref{fansimple})  and (\ref{qhalf}) in the case of $e_1=1$:
\begin{eqnarray*}
H(1, e_2, \ldots ,e_d) & = & \int_0^\infty   - \ln \left[\frac{\tau^d}{q(\tau)}\right]  + \ln \left[\frac{\tau}{\tau+1} \right] \,\, d\tau  \\
Q(1, e_2, \ldots ,e_d) & = & \int_0^\infty \hspace{4mm} -\left[\frac{\tau^d}{q(\tau)}\right]  + 
\left[ \frac{\tau}{\tau+1}\right] \,\, d\tau
\end{eqnarray*}

The formulae in eqs. (\ref{hhalf}) and  (\ref{qhalf}) immediately yield half-axis
formulae for partial derivatives of $H$ and $Q$ with respect to the
symmetric polynomials. For instance
\begin{equation}\label{halfintk}
  \pder{H}{e_k}= \int_0^\infty \frac{\tau^{d-k}}{q(\tau)}\,\, d\tau,
  \hspace{3mm}\mbox{for $k\geq 2$} \end{equation}
and
\begin{equation}\label{halfint1}
  \pder{H}{e_1}=  \int_0^\infty \left[ \, \frac{\tau^{d-1}}{q(\tau)}-\frac{1}{(\tau+1)}\, \right]\, d\tau\,\, -1
  \hspace{3mm}\mbox{for $k=1$} .\end{equation}

\section{Properties of H and Q}\label{consequences} 

A variety of properties of $H$ and $Q$ follow from their half-axis
formulae given above. We describe these in full in the Appendix, but we summarise here a few key results, pointing the reader to the Appendix for further details and proofs. Some of these results will be used in our developments in the next section.

In the next section we will build on the fact that the partial
derivaties of $H$ and $Q$ have definite signs:
\begin{align}
\label{Hone} &\pder{H}{e_k} > 0, \hspace{1cm} 2 \le k \le d,\\
\label{Qone} &\pder{Q}{e_k} >0, \hspace{1cm} \mbox{ for } 2 \le k \le d,\\
\label{Hhigher} (-1)^{m-1} &\frac{\partial^m H}{\partial e_{k_1}\ldots \partial e_{k_m}} > 0, \hspace{1cm}  \mbox{ for } m \ge 2, \ \ 1 \le k_j \le d.\\
\label{Qhigher} (-1)^{m-1} &\frac{\partial^m Q}{\partial e_{k_1}\ldots \partial e_{k_m}} >0, \hspace{1cm} \mbox{ for } m \ge 2, \ \ 1 \le k_j \le d.
\end{align}
See Appendix, Property \ref{prop4}. Note that $\partial H/\partial e_1$ and $\partial Q/\partial e_1$  can be negative, so the restriction to $k \ge 2$ in inequalities (\ref{Hone}) and (\ref{Qone}) is necessary. However the indices of the $e$'s in inequalities (\ref{Hhigher}) and (\ref{Qhigher}) can lie in the whole range $1, \ldots, d$. 

We will also use the fact that the $m^{\rm th}$ derivatives of $H$ and $Q$ depend
only on the {\em sum} of the indices, $\sum k_j$. Thus for example
$\partial^2 H/\partial e_1 \partial e_5 = \partial^2 H/\partial
e_2 \partial e_4=\partial^2 H/\partial e_3^2$ since $1+5=2+4=3+3$. See
Property \ref{indices}.

A surprisingly close functional relationship between $H$ and $Q$, not so apparent from the defining eqs. (\ref{heq}) and (\ref{qdeq}), is indicated in the equalities
\begin{align*}
H&=e_1+\sum_{k=1}^d ke_k\pder{H}{e_k};\\
Q&=e_1+\sum_{k=1}^d e_k\pder{H}{e_k},
\end{align*}
and
\begin{align*}
-\pder{Q}{e_k}=\frac{\partial^2 H}{\partial e_l \,\partial e_m},
\end{align*}
for any $k,l,m$ with $k=l+m$ and $l,m\geq 1$. See Appendix, Properties \ref{sumprop} and \ref{QHder}.

One can also obtain some bounds, for instance
\begin{align*}
H-Q \ge \sum_{k=2}^d \frac{d^{k-1}e_k}{\binom{d-1}{k-1}e_1^{k-1}}.
\end{align*}
See Property \ref{difference}. If the $e_k$'s are the symmetric
polynomials in the eigenvalues of a state $\rho$, then this is a lower
bound on the range of the accessible information obtainable from pure
state ensembles with density matrix $\rho$. Even the first term is
meaningful, for it says
\[
H-Q \ge \frac{d e_2}{(d-1)e_1},
\]
implying that the range is non-zero when the state $\rho$ is not a
pure state. 

Finally, we mention the following upper bound for $H$ (there is a
corresponding one for $Q$), which depends only on $e_1$ and $e_2$:
\begin{align*}
H(e_1,e_2, \ldots, e_d) \le -(d-1)a\ln a-b\ln b,
\end{align*}
where $a$ and $b$ are roots of $(n-1)a+b=e_1$ and ${n-1 \choose
  2}a^2+(n-1)ab=e_2$. See Property \ref{HQbounds}. A similar type of
bound (i.e. depending only on $e_1$ and $e_2$) was obtained by
Hellmund and Uhlmann \cite{HU}. However, the above bound is optimal amongst all bounds that depend only on $e_1$ and $e_2$, as 
it is always attained by an actual assignment $(x_1, \ldots ,x_d) = (a, \ldots ,a,b)$ satisfying the given $e_1$ and $e_2$ values.

\section{H and Q as Bernstein and Pick functions}\label{picky}

A function $f(s)$ is said to be {\em completely monotone} if $(-1)^k
d^kf/ds^k \ge 0$, for $s \in [0,\infty)$ and $k\ge 0$. We extend this definition to several variables by requiring that
\begin{align}
\label{cm}  (-1)^k \partial^k f/\partial s_{i_i} \ldots \partial s_{i_k} \ge 0,\ \ \mbox{for   } \ \ s_{i_j} \ge 0, \ \ k \ge 0.
\end{align} 
Bernstein's theorem \cite{bernsteinpaper,feller,Bernstein-functions} asserts that a function $f(s)$ of one variable is completely monotone iff $f(s)$ is the
Laplace transform of a non-negative density $\nu$,
viz. \[f(s)= \int_0^\infty e^{-st}\nu (t) \, dt. \]
Here the density $\nu$ is generally supported on the half {\em closed} interval $[0,\infty )$ \cite{Bernstein-functions} and may involve point weights (e.g. the completely monotone $f(s)=1$ has $\nu (t)=\delta (t)$).
If $b$ is the weight of the point $t=0$ then the Bernstein representation may also be wirtten as 
\[ f(s)=b+\int_{0^+}^\infty e^{-st}\nu (t) \, dt. \]
In our applications we will always have $b=0$.

  We can generalise Bernstein's representation theorem  to multivariate completely monotone functions as follows.

\begin{lemma}\label{berlemma}
$f(s_1, \ldots ,s_m)$ is completely monotone iff 
$f(s_1, \ldots,
s_m)=\cL[\xi(t_1, \ldots, t_m)]$ where $\xi$ is a non-negative density for $t_i \ge 0$, $1 \le i \le m$. Explicitly the condition is
\begin{align}\label{multilaplace}
f(s_1, \ldots, s_m)=\int_0^\infty e^{-(s_1t_1 + \ldots +s_mt_m)} \xi(t_1, \ldots ,t_m) \, dt_1 \ldots dt_m.
\end{align}

\end{lemma}
\noindent {\bf Proof}\, Consider first the case of two variables.  Since $f(s_1,s_2)$ is completely
monotone in the first variable, Bernstein's theorem implies that
$f(s_1,s_2)=\cL[\nu(t_1,s_2)]$, the Laplace transform applying only to
the first variable. Next note that eq. (\ref{cm}) implies that for each $k$, $(-1)^k\partial^k f/\partial s_2^k$ is completely monotone in $s_1$ so it is the Laplace transform of a non-negative function. But we already have 
$(-1)^k\partial^k f/\partial s_2^k
=\cL[(-1)^k\partial^k \nu(t_1,s_2)/\partial s_2^k ]$ so by uniqueness of the Laplace transform, 
$(-1)^k\partial^k \nu(t_1,s_2)/\partial s_2^k $ must be non-negative for all $s_2\geq 0$. Thus $\nu (t_1,s_2)$ is completely monotone in $s_2$ and so itself must be the Laplace transform (for the single variables $t_2$ and $s_2$) of some non-negative $\xi(t_1,t_2)$. Finally then 
$f(s_1,s_2)=\cL[\nu(t_1,s_2)]=\cL[\xi(t_1,t_2)]$. This argument
readily extends to any number of variables.\,\, $\Box$

Equations (\ref{Hone}) - (\ref{Qhigher}) (see Property \ref{prop4})  can be interpreted as asserting that each first
derivative $\partial H/\partial e_k$ and $\partial Q/\partial e_k$,
for $2 \le k \le d$, is completely monotone in the multivariate
sense. But what about $H$ and $Q$ themselves?
Note first that the variable $e_1$ has a distinguished role since $\partial H/ \partial e_1$ and $\partial Q/partial e_1$ are not generally non-negative. Furthermore $H$ itself is not generally non-negative for $e_1>1$ (e.g. recall that $H(x_1, \ldots ,x_d)=-\sum x_i \ln x_i$ and consider $e_k$'s arising from a large positive $x_1$ and suitably small positive $x_2,\ldots , x_d$). Thus let us set $e_1=1$ and introduce the 
positive cone $\cE^+$
defined by $e_k \ge 0$, $2 \le k \le d$  (though, as pointed in the
Introduction, only part of this cone corresponds to real, positive
$x_i$).  We know $H=0$ when $e_1=1$ and $e_2= \ldots =e_d=0$, since
this corresponds to one of the underlying probabilities being 1 and
the others 0. But then every point in $\cE^+$ can be reached by moving
along its coordinate axes independently, and it follows from $\partial
H/\partial e_k > 0$ for $k\geq 2$, that $H$ must be positive everywhere in
$\cE^+$. By Property \ref{QHder}, a similar conclusion applies to $Q$.

Although non-negative, $H(1, e_2, \ldots , e_d)$ is not completely monotone, since both it and its first
derivatives are non-negative: there is no change of sign between the
function and its first derivative, as the definition
requires. Similarly, $ Q(1, e_2, \ldots ,e_d)$ is not completely monotone. However, a
function that is non-negative on the positive cone and has completely
monotone first derivatives is said to be a {\em Bernstein}
function and from Property \ref{prop4} we have
\begin{proposition}
$H(1, e_2, \ldots , e_d)$ and $Q(1, e_2, \ldots , e_d)$ are multivariate Bernstein functions.
\end{proposition} 
\noindent There is an extensive literature on Bernstein functions
\cite{Bernstein-functions}. Any such function has a {\em
  Levy-Khintchine representation}, which in one variable takes the
form 
\begin{align}
\label{LK} f(s)=a+bs+\int_0^\infty(1-e^{-st})\mu(t) \, dt,
\end{align}
for some non-negative $\mu$ supported on $(0,\infty )$, satisfying $\int_0^\infty
\min(1,t)\mu(t)\, dt < \infty$. This can be obtained \cite{Bernstein-functions} by integrating
Bernstein's Laplace transform representation of $f'(s)$ mentioned
above. The constants $a$ and $b$ can be identified \cite{Bernstein-functions} as $a=f(0)$ and $b=\lim_{s\rightarrow \infty} f(s)/s$. In our applications and extensions below we will always have $a=b=0$.

As with Lemma \ref{berlemma} and Bernstein's theorem, the Levy-Khintchine representation may also be extended to multi-variate functions.
We illustrate this by deriving the
Levy-Khintchine representation for $H(1, e_2, \ldots ,
e_d)$. Recollecting eq. (\ref{halfintk}), we have
\[
\frac{\partial H(1,e_2, \ldots, e_d)}{\partial
e_k}= \int_0^\infty \frac{\tau^{d-k}}{\tau^d+\tau^{d-1}+e_2\tau^{d-2} + \ldots e_d} \, d\tau.
\]
But we can write the integrand for any given $\tau$ as a Laplace
transform of the variable $e_d$, i.e. assigning $e_d$ the role of
`$s_d$' in eq. (\ref{multilaplace}):
\[
\frac{\tau^{d-k}}{\tau^d+\tau^{d-1}+e_2\tau^{d-2} + \ldots e_d}=\int_0^\infty \tau^{d-k} e^{-e_dt_d}\, e^{-(\tau^d+\tau^{d-1}+ \ldots + e_{d-1}\tau)t_d} \,\,dt_d.
\]
Next we write the integrand above as the Laplace transform with respect to
the variable $e_{d-1}$ to give the double transform:
\begin{align*}
\frac{\tau^{d-k}}{\tau^d+\tau^{d-1}+e_2\tau^{d-2} + \ldots e_d}&=\\
\int_0^\infty \tau^{d-k} e^{-(e_{d-1}t_{d-1}+e_dt_d)}\,  &e^{-(\tau^d+\tau^{d-1}+ \ldots e_{d-2}\tau^2)t_d} \, \delta \left(t_{d-1}-\tau t_d \right) dt_{d-1}\, dt_d.
\end{align*}
Continuing in this way we find
\begin{align*}
\frac{\tau^{d-k}}{\tau^d+\tau^{d-1}+ \ldots e_d}=
\int_0^\infty \tau^{d-k} e^{-\sum_{i=2}^d e_it_i}\,  e^{-(\tau^d+\tau^{d-1})t_d} \prod_{i=1}^{d-2}\delta \left(t_{d-i}-\tau^i t_d \right) dt_2 \ldots dt_d.
\end{align*}
Integrating over $\tau$ now gives
\begin{align}\label{Lpartials}
\frac{\partial H(1,e_2, \ldots, e_d)}{\partial
e_k}=\cL[\phi_k(t_2,\ldots ,t_d)],
\end{align}
where
\begin{align}\label{phi}
\phi_k(t_2, \ldots, t_d)=\int_0^\infty  \tau^{d-k} e^{-(\tau^d+\tau^{d-1})t_d} \prod_{i=1}^{d-2}\delta \left(t_{d-i}-\tau^i t_d \right) d\tau.
\end{align}
Now let us pause and consider the consequences of the Laplace
transforms (\ref{Lpartials}). We can differentiate under the integral to obtain
\begin{align}
\frac{\partial^2 H(1,e_2, \ldots, e_d)}{\partial e_j \partial
e_k}=\cL[t_j\phi_k(t_2,\ldots ,t_d)]=\cL[t_k\phi_j(t_2,\ldots ,t_d)],
\end{align}
which, by the uniqueness of the Laplace transform, implies
\begin{align}\label{phi-relations}
t_j\phi_k(t_2,\ldots ,t_d)=t_k\phi_j(t_2,\ldots ,t_d),
\end{align}
for $2 \le j,k \le d$. This means we can define a function
$\mu$ by
\begin{align}\label{mu}
\mu(t_2, \ldots, t_d)=\frac{\phi_2(t_2, \ldots , t_d)}{t_2}=\ldots =\frac{\phi_d(t_2, \ldots , t_d)}{t_d}.
\end{align}
Now let us integrate the equations (\ref{Lpartials}) to obtain
\begin{align*}
H(1,e_2,0,\ldots,0)-H(1,0,0,\ldots 0)&=\int_0^\infty \frac{1-e^{-e_2t_2}}{t_2}\phi_2(t_2,\ldots,t_d)dt_2 \ldots dt_d\\
&=\int_0^\infty \left(1-e^{-e_2t_2}\right)\mu(t_2,\ldots,t_d)dt_2 \ldots dt_d,\\
H(1,e_2,e_3,\ldots,0)-H(1,e_2,0,\ldots 0)&=\int_0^\infty e^{-e_2t_2}\frac{1-e^{-e_3t_3}}{t_3}\phi_3,\\
&=\int_0^\infty \left(e^{-e_2t_2} - e^{-(e_2t_2+e_3t_3)}\right)\mu(t_2,\ldots,t_d)dt_2 \ldots dt_d.
\end{align*}
Continuing in this way to $H(1, \ldots ,e_{d-1},e_d)-H(1, \ldots ,e_{d-1},0)$, then adding all the equations, and using $H(1,0,0,\ldots 0)=0$, we get
\begin{align}\label{multiLK}
H(1,e_2,e_3,\ldots,e_d)=\int_0^\infty \left(1 - e^{-\sum_{i=2}^d e_i t_i}\right)\mu(t_2,\ldots,t_d)\, dt_2 \ldots dt_d.
\end{align}
If we evaluate $\phi_k$ from eq. (\ref{phi}) we can use eq. (\ref{mu})
to give $\mu$ explicitly. Let us first, however, make a further
deduction from the Laplace transforms (\ref{Lpartials}). Property
\ref{indices} says that higher derivatives depend only on the sums of
indices. This implies further relations besides
eq. (\ref{phi-relations}), and these entail relations between the
variables $t_i$. For example, $\partial^2 H/\partial e_2 \partial e_5
= \partial^2 H/\partial e_3 \partial e_4$ implies
$t_2\phi_5=t_3\phi_4$, which eq. (\ref{mu}) allows us to write as
$t_2t_5\mu=t_3t_4 \mu$, or $t_2t_5=t_3t_4$. This tells us that the
density $\mu$ of the Levy-Khintchine representation we seek lies in
some variety within the positive cone $\RR^{d-2}_+$ defined by the $t_i$,
and we can see this also from the product of delta-functions in
eq. (\ref{phi}), since this product is only non-vanishing when there
is a value of $\tau$ which simultaneously satisfies $t_{d-i}=\tau^i
t_d$ for $1 \le i \le d-2$. We now characterise that variety:
\begin{lemma}\label{algebraic}
  Property \ref{indices} implies $d-3$ independent relations amongst
  the $d-1$ variables $t_2, \ldots , t_d$, namely
\begin{align}\label{relations}
t_dt_i=t_{d-1} t_{i+1} \mbox{\hspace{2mm} for \hspace{2mm} } 2 \le i \le d-2.
\end{align}
These are also the relations obtained by eliminating $\tau$ from
\begin{align}\label{tau}
t_{d-i}=\tau^i t_d  \mbox{\hspace{2mm} for\hspace{2mm} } 1 \le i \le d-2,
\end{align}
which must be satisfied for the non-vanishing of the product of
delta-functions in eq. (\ref{phi}). These $d-3$ relations define a 2D
surface in the space of the $t_i$.
\end{lemma}
\noindent \begin{proof}
The relations following from Property \ref{indices} have the form $t_it_j=t_kt_l$, for $i+j=k+l$. To show that any such relation can be obtained from eqs. (\ref{relations}), note that we can combine two of them to get $(t_dt_i)(t_{d-1}t_j)=(t_{d-1}t_{i+1})(t_dt_{j+1})$, which implies $t_it_j=t_{i+1}t_{j-1}$. Iterating gives $t_it_j=t_{i+1}t_{j-1}=t_{i+2}t_{j-2}= \ldots =t_kt_{j-(k-i)}$, which is a general relation of the form we seek. It is easy to see that eqs. (\ref{relations}) are equivalent to eqs. (\ref{tau}) with  $\tau=t_{d-1}/t_d$.\,\, $\Box$
\end{proof}
\begin{theorem}
  The density $\mu$ in eq. (\ref{multiLK}), the Levy-Khintchine representation of $H(1,e_2,
  \ldots, e_d)$, is given by
\begin{align}\label{mu-theorem}
\mu(t_2, \ldots, t_d)=\frac{1}{t_d^2}\exp \left(-t_d\left(r^d+r^{d-1}\right)\right)\prod_{i=2}^{d-2} \delta(t_i-rt_{i+1})
\end{align}
  where $r=t_{d-1}/t_d$. The corresponding density for  $Q(1,e_2,
  \ldots, e_d)$, is given by
\begin{align}
\mu(t_2, \ldots, t_d)=\frac{r^d}{t_d}\exp \left(-t_d\left(r^d+r^{d-1}\right)\right)\prod_{i=2}^{d-2} \delta(t_i-rt_{i+1})
\end{align}
 The product of delta-functions restricts the
function to the surface defined by the relations eq. (\ref{relations}) in
Lemma \ref{algebraic}.
\end{theorem}

\noindent\begin{proof} Write 
\[
\alpha(t_2, \ldots, t_d)=\left(1 - e^{-\sum_{i=2}^d e_i t_i}\right)
\]
and 
\[
\beta(t_{d-1},t_d)=\frac{1}{t_d^2}\exp \left(-t_d\left((t_{d-1}/t_d)^d+(t_{d-1}/t_d)^{d-1}\right)\right).
\]
If we define $\mu$ by eqs.  (\ref{mu}) and (\ref{phi}) with $k=d$ and substitute it into  eq. (\ref{multiLK}), we get
\begin{align}\label{version2}
H(1,e_2, \ldots, e_d)=\int_0^\infty \alpha(\tau^{d-2}t_{d}, \ldots, \tau t_{d},t_d) \beta(\tau t_d,t_d) t_d\, d\tau\, dt_d.
\end{align}
Then replacing the variable $\tau$ by $t_{d-1}$ via $\tau= t_{d-1}/t_d$, we get
\begin{align}\label{version1}
H(1,e_2, \ldots, e_d)=\int_0^\infty \alpha((t_{d-1}/t_d)^{d-3}t_{d-1}, \ldots, (t_{d-1}/t_d)t_{d-1}, t_{d-1},t_d) \beta(t_{d-1},t_d) \, dt_{d-1}\, dt_d
\end{align}
which is equivalent to inserting the claimed formula eq. (\ref{mu-theorem}) for $\mu$  into eq. (\ref{multiLK}), and integrating out the delta functions.

For $Q$ we follow a similar argument at all the steps, starting from 
\[
\frac{\partial Q}{\partial e_k}=\int_0^\infty \frac{\tau^{2d-k}}{q(\tau)^2} d \tau,
\]
and using $1/(s+a)^2=\cL \left[ te^{-at} \right]$.
\,\, $\Box$
\end{proof}

$H(1, e_2, \ldots , e_d)$ and $Q(1, e_2,
\ldots ,e_d)$, regarded as functions of each single $e_k$ separately,
can be seen to belong to a further special class, the {\em complete
  Bernstein functions}, where $\mu$ in eq. (\ref{LK}) (for functions of one variable) is itself
completely monotone. Equivalently \cite{Bernstein-functions}, such
functions can be characterised as non-negative valued functions that
are operator monotone functions, or Pick functions. A function $f$ is
operator monotone if $A \le B$ implies $f(A) \le f(B)$ for any
self-adjoint matrices $A$, $B$. A Pick function is a real-valued
function on $[0,\infty)$ that possesses an analytic continuation on
$\complex\backslash(-\infty,0)$ that maps the upper half plane into
itself. It is known \cite{bhatia} that $f$ is operator monotone iff
$f$ is a Pick function.

\begin{proposition}\label{pickHQ}
$H(1, e_2, \ldots , e_d)$ and $Q(1, e_2, \ldots , e_d)$ are Pick functions in each of their variables separately (with all other variables being set to any fixed positive real values).
\end{proposition} 
\noindent {\bf Proof}\, For the case of $H$, setting $e_1=1$ in eq. (\ref{hhalf}) we see that the integrand has the form $r_1(\tau )/r_2(\tau )$ with $r_2(\tau)=q(\tau)(\tau +1)$ having all roots on the negative real axis and $r_1$ being of degree two less than $r_2$ for all complex values of $e_2, \ldots ,e_d$. Hence if any $e_k$ is extended to $\complex\backslash (0,\infty )$, the integral remains finite and is a holomorphic function of the chosen $e_k$. Furthermore, setting $e_1=1$ in the alternative formula eq. (\ref{fansimple})  (and all other variables apart from $e_k$ being set to positive real values) we get
\[ H (1,e_2, \ldots , e_d)=\int_0^\infty
\, [ \, \ln (\tau^d+\tau^{d-1}+e_2\tau^{d-2}+\ldots +e_d)+(1-d)\ln \tau-\ln (\tau+1)\,]\, d\tau. \]
so the upper half plane will be preserved by $H$, since if $\Im(z)$ denotes the imaginary part of $z$, we have $\Im(z) >
0$ implies $\arg(z) > 0$ i.e. $\Im( \ln z) > 0$ and $\Im( H) > 0$.\\
For the case of $Q$, setting $e_1=1$ in eq. (\ref{qhalf}) we get
\[ Q (1,e_2, \ldots , e_d)=\int_0^\infty
\, \left[ \, \frac{-\tau^d}{ (\tau^d+\tau^{d-1}+e_2\tau^{d-2}+\ldots +e_d)}-\frac{1}{(\tau+1)}  +1 \,\right] \, d\tau. \]
Then extending $e_k$ to the complex plane (with other variables remaining real), the imaginary part of the integrand will be $\Im(e_k)\tau^{2d-k}/|q|^2$ so the upper half plane will be preserved by $Q$.\,\, $\Box$

There is yet another way of looking at Property \ref{prop4}. If $f$ is
a function whose first derivatives are completely monotone, then
$e^{-f}$ is completely monotone \cite{feller}. This is easy to check
by repeated differentiation of $e^{-f}$. It follows that $e^{-H(1,e_2,
  \ldots, e_d)}$ is the Laplace transform of a non-negative function
$\mu(t_2, \ldots, t_d)$, and since $H(1,0, \ldots, 0)=0$, $\mu$ is a
probability density. Actually, we can say more than this, since
$e^{-H(1, e_2, \ldots , e_d)}=(e^{-H/m})^m=\cL[\nu^{*m}]$, where $e^{
  -H/m}=\cL[\nu]$. This means that, for any integer $m$, $\mu$ is the
$m$-fold convolution of a measure $\nu$. This property is called {\em
  infinitely divisibility} \cite{feller}, and is possessed by many
fundamental statistical distributions, such as the Gaussian. Thus we
know that $e^{-H(1, e_2, \ldots , e_d)}$ is the Laplace transform of
an infinitely divisible function, and, since all the above remarks
apply to $Q$, the same is true of $e^{-Q(1, e_2, \ldots , e_d)}$.

Finally, we observe that the Levy-Khintchine representations for $H$ and $Q$ are multivariate versions of eq. (\ref{LK}) where the constants $a$ and $b$ are both zero. They are therefore what Audenaert \cite{koenraad} calls {\em bare} Bernstein functions. He shows (in the single variable case) that such functions satisfy various properties, including a type of matrix trace inequality. It would be interesting to know if this result, and also the operator monotone property, can be extended to the multivariate case.  Ways of extending the concept of operator monotone functions to many variables have been explored \cite{Agler,Hansen}.

\section{Concluding remarks}
We have noted in the Introduction that informational properties of a random variable should be independent of the labelling of its outcomes and hence be represented by symmetric functions of the probabilities ($x_j$'s). Then on passing to the elementary symmetric polynomials ($e_k$'s) as variables, we have seen that the resulting functional forms of the entropy and subentropy exhibit a series of novel structural properties. A further benefit of the $e_k$'s is that derivatives and integrals with respect to these variables automatically preserve symmetry with respect to the $x_j$'s (in contrast to say $\partial f(x_1, \ldots ,x_d)/\partial x_j$ for a symmetric function $f$) and hence we have the benefit of the full power of calculus within the ``informationally meaningful'' regime of constructions that are symmetric in the probabilities.

A further intriguing feature of the $e_k$ variables is that the entropy and subentropy become functionally surprisingly closely related. As subentropy is an intrinsically {\em quantum} informational construct (whereas the entropy function features fundamentally in both classical and quantum information theory) this suggests that the $e_k$ variables may offer a special advantage for studying the extra intricacies of quantum over classical information theory. Such an advantage was also suggested in \cite{mj04} where the monotonicity property of the first derivatives $\partial H/\partial e_k$ for $k\geq 2$ was used to provide a new interpretation of Schumacher compression of quantum information (which reduced to a trivial statement for the sub-case of classical information compression). But subentropy aside, the novel properties we have seen for $H(e_1, \ldots , e_d)$ itself suggest a role for the $e_k$ variables already in just classical information theory.

Complex numbers have repeatedly played a key role in our constructions. In fact, they arise in three distinct ways:

Firstly, although non-negative $x_j$'s map to non-negative $e_k$'s, the image of this map is only a subset of $\RR^d_+$, and it is natural to analytically extend the entropy functions to the full positive cone $\RR^d_+$ of all $d$-tuples of non-negative $e_k$'s. But then, mapping back to the space of $x_j$'s amounts to allowing probabilities to become complex, occurring always as complex conjugate pairs, although never being real and negative. 

Secondly, we note that our half-axis formulae for entropy and subentropy were transparently derived starting from complex contour integral expressions, exploiting the basic relation eq. (\ref{polypeq}) between the $x_j$'s and $e_k$'s, together with properties of the complex logarithm function. In standard information theory and thermodynamics, the logarithm function serves to endow entropy with a fundamental extensionality property for composite independent systems, via $\ln ab = \ln a+\ln b$, whereas for us (cf the key-hole countour integral)  it plays an entirely different role: the {\em complex} logarithm function serves to provide a constant discontinuity of $2\pi$ in its imaginary part along its branch cut $(-\infty, 0)$. 

Thirdly, complex numbers feature again in the notion of Pick functions and the mapping properties of $H(1,e_2,\ldots ,e_d)$ and  $Q(1,e_2,\ldots ,e_d)$, when each $e_k$ for $k\geq 2$ is allowed to vary over the upper half complex plane. 

All of the above suggests that some features of information theory may acquire a special simplicity if formulated in an enlarged mathematical setting that includes the elementary symmetric polynomials and complex variables. Yet, paradoxically, there is no obvious probabilistic interpretation of the underlying ingredients. Extending probabilities $x_k$ into the complex domain has no immediately apparent meaning; nor can one interpret the elementary symmetric polynomials as probabilistic objects, except perhaps in a trivial way,  e.g. regarding $k!e_k$ as the probability that $k$ samples of a random variable with probabilities $\{ x_1, \ldots ,x_d\}$ yield no repeated outcome. This is not the sort of meaning we seek, and discovering this elusive deeper meaning is a fascinating puzzle.


\section{Appendix:  properties of $H$ and $Q$}

Here we derive various relations and inequalities from the half-axis formulae. Any proofs that are omitted follow straightforwardly from the half-axis formulae, and are left to the reader to provide.
First, we have
\begin{property}\label{sumprop}
\begin{align}
\label{hsum}H&=e_1+\sum_{k=1}^d ke_k\pder{H}{e_k};\\
\label{qsum}Q&=e_1+\sum_{k=1}^d e_k\pder{H}{e_k}.
\end{align}
\hspace{14cm} $\Box$
\end{property}
It is striking how similar $H$ and $Q$ appear in this formulation,
compared to the very different-looking expressions eq. (\ref{heq}) and
eq. (\ref{qdeq}). We can make eq. (\ref{hsum}) look even closer to
eq. (\ref{qsum}) by introducing the variables $f_k=e_k^{1/k}$ to get
\begin{align}
H=f_1+ \sum_{k=1}^d f_k  \pder{H}{f_k}.
\end{align}
Note that there is an analogous expression for $H$ in terms
of derivatives with respect to the $x_k$, obtained directly from
eq. (\ref{heq}):
\begin{align}
\label{xsum}H=e_1+ \sum_{k=1}^d x_k  \pder{H}{x_k}.
\end{align}

Since the integrand in eq. (\ref{halfintk}) is positive, it follows
that $\partial H/\partial e_k \ge 0$ for $k \ge 2$
\cite{fannes}. However, one can show something a little stronger,
since
\[
\int_0^\infty \frac{\tau^{d-k}}{q} d\tau \ge \int_o^\infty \frac{\tau^{d-k}}{(\tau+e_1/d)^d} d\tau = c_{d,k},
\]
where 
\begin{align}\label{c}
c_{d,k}=\frac{d^{k-1}}{(d-k+1)\binom{d-1}{k-2}e_1^{k-1}}.
\end{align}
Thus we get the positive lower bound
\begin{property} \label{oneder} 
\begin{align}
\pder{H}{e_k} \geq c_{d,k} > 0 \hspace{3mm} \mbox{for $k\geq 2$.} 
\end{align}
\hspace{14cm} $\Box$
\end{property}
This was proved in \cite{mj04}, by a more complicated
argument. Combining this with eq. (\ref{hsum}) and eq. (\ref{qsum}), we get
\[
H-Q=\sum_{k=2}^d (k-1)e_k\pder{H}{e_k} \ge \sum_{k=2}^d \frac{d^{k-1}(k-1)e_k}{(d-k+1)\binom{d-1}{k-2}e_1^{k-1}},
\]
or 
\begin{property}\label{difference}
\begin{align}
H-Q \ge \sum_{k=2}^d \frac{d^{k-1}e_k}{\binom{d-1}{k-1}e_1^{k-1}}.
\end{align}
\hspace{14cm} $\Box$
\end{property}

One also has:
\begin{align}
\label{oneder1} \pder{H}{e_1} \leq -1 \hspace{3mm} \mbox{if $e_1\geq 1$.} 
\end{align}

A further type of connection between $H$ and $Q$ is given by
\begin{property}\label{QHder}
\begin{align}\label{qder1}
-\pder{Q}{e_k}=\frac{\partial^2 H}{\partial e_l \,\partial e_m}\hspace{3mm}\mbox{for any $k,l,m$ with $k=l+m$ and $l,m\geq 1$.}
\end{align}
\hspace{14cm} $\Box$
\end{property}
Note that this does not arise directly from  differentiation of eq. (\ref{qsum}).

\begin{property}\label{indices}
The $m^{\rm th}$ derivative $ \partial^m H/ \partial e_{i_1}\ldots \partial e_{i_m}$  as a function of $(e_1, \ldots ,e_d )$ depends only on the sum of indices $i_1+\ldots +i_m$, and the same holds for $Q(e_1, \ldots ,e_d)$ too.\,\, $\Box$

\end{property}

\begin{property} \label{reduce} Consider the $m^{\rm th}$ derivative $\partial^m H/\partial e_{i_1}\ldots \partial e_{i_m}$ for $H(e_1, \ldots ,e_d)$ with $d$ variables. Introduce the entropy function $\tilde{H}(\tilde{x}_1, \ldots , \tilde{x}_{dm})$ with $dm$ variables and corresponding elementary symmetric polynomials $\tilde{e}_1, \ldots , \tilde{e}_{dm}$. Then for any $(e_1, \ldots , e_d)$ arising from roots $x_1, \ldots x_d$, any $m^{\rm th}$ derivative of $H$ can be expressed as a {\em first} derivative of $\tilde{H}$:
\begin{equation}(-1)^{m-1}\frac{\partial^m H}{\partial e_{i_1} \ldots \partial e_{i_m}} (e_1, \ldots ,e_d) =
\pder{\tilde{H}}{\tilde{e}_{K}}(\tilde{e}_1, \ldots , \tilde{e}_{md}) \end{equation}
where $K=i_1+\ldots +i_m$ and the RHS is evaluated at the point $(\tilde{e}_1, \ldots , \tilde{e}_{md})$ being the elementary symmetric polynomial values for the $md$  $\tilde{x}_j$'s 
\[  (\tilde{x}_1, \ldots , \tilde{x}_{dm}) = (x_1, \ldots , x_1, x_2, \ldots ,x_2, \hspace{2mm} \ldots\hspace{2mm} , x_d, \ldots ,x_d)\]
having each $x_i$ repeated $m$ times.\,\, $\Box$
\end{property}


Each time we differentiate eq. (\ref{halfintk}) there is a switch in
sign. Combining this with Properties \ref{oneder}, \ref{reduce} and
\ref{QHder} gives:
\begin{property}\label{prop4}
For $m\geq 2$ we have
\begin{align} 
(-1)^{m-1} \frac{\partial^m H}{\partial e_{i_1}\ldots \partial e_{i_m}} &\geq c_{md,i} > 0\hspace{3mm} \mbox{for all $i_1,\ldots ,i_m \geq 1$,}\\
(-1)^{m-1} \frac{\partial^m Q}{\partial e_{i_1}\ldots \partial e_{i_m}} &\geq  c_{(m+1)d,i} >0\hspace{3mm} \mbox{for all $i_1,\ldots ,i_m \geq 1$,}
\end{align}
where $i=\sum_l i_l$ and $c_{d,k}$ is given by eq. (\ref{c}). For first derivatives we have \[
\pder{H}{e_i}\geq c_{d,i} > 0 \mbox{ and } \pder{Q}{e_i}\geq c_{2d,i}>0 \mbox{ for }
i\geq 2.
\] \hspace{14cm} $\Box$
\end{property}
Note that the above now subsumes Property \ref{oneder}.


\begin{property}
For  $\theta \ge 1$
\begin{align}
\label{Qx} Q(\theta x) &\le \theta Q(x),\\
\label{Qe} Q(\theta e) &\le \theta Q(e),\\
\label{Hx} H(\theta x) &\le \theta H(x),\\
\label{He} H(\theta e) &\ge \theta H(e),\\
\label{HQ} Q(\theta x)-\theta Q(x)&=H(\theta x)-\theta H(x)
\end{align}
where $Q(x)$ and $Q(e)$ denote the subentropy $Q$ regarded as a function of the $x_k$'s and $e_k$'s respectively
(and similarly for $H$).\,\, $\Box$
\end{property}
\noindent{\bf Proof}\, 
By definition
\begin{align*}
\frac{1}{\theta}Q(\theta x)&=-\sum_k \frac{x^d_k \ln(\theta x_k)}{\prod_{i \ne k} (x_k-x_i)}
=Q(x)-\sum_k \frac{x^d_k}{\prod_{i \ne k} (x_k-x_i)}\ln\theta\\
&=Q(x)-e_1 \ln\theta,
\end{align*}
(where the last equality follows since $\sum_k \frac{x^d_k}{\prod_{i \ne k} (x_k-x_i)}$ is a rational symmetric function, homogeneous of degree one, with all singularities removable, so it must be a multiple of $e_1$.)
Since $e_1\ln\theta\ge 0$ for $\theta \ge 1$, this proves
inequality (\ref{Qx}). Inequality (\ref{Hx}) is immediate from eq. (\ref{heq}), and eq. (\ref{HQ}) follows.

To prove inequality (\ref{Qe}), introduce $\xi=q(\tau)-\tau^d=e_1\tau^{d-1}+e_2\tau^{d-2}+\ldots +e_d$ and then eq. (\ref{qhalf}) directly gives
\[ \theta Q(e)-Q(\theta e)=\int_0^\infty \theta\left[  -\frac{\tau^d}{\tau^d+\xi}+1 \right] -
\left[ -\frac{\tau^d}{\tau^d+\theta\xi}+1\right] \, d\tau .\]
 After a little algebra, the integrand simplifies to $\frac{(\theta-1)\theta\xi^2}{q(\tau)(\tau^d+\theta\xi)}$
which is positive for all $\tau>0$ and $\theta>1$, and inequality (\ref{Qe}) follows. Similarly, from eq. (\ref{hhalf}) we obtain $\theta H(e)-H(\theta e) = \theta(\theta-1)\int_0^\infty \frac{\xi(\tau q^\prime(\tau)-dq(\tau))}{q(\tau)(\tau^d+\theta\xi)}$, which is negative for all  $\tau>0$ and $\theta>1$ since $\tau q^\prime(\tau)-dq(\tau) \le 0$. This gives inequality (\ref{He}).
\,\, $\Box $

Our next result applies to bipartite systems. Although the linear operation of forming marginals of a joint probability distribution does not translate naturally into an operation on the corresponding symmetric polynomials, we still have the following result in that context.

Let $\RR^n_+=\{ (\tau_1, \ldots ,\tau_n):\tau_i \geq 0,\,\, i=1, \ldots ,n \}$ be the positive cone for $n$ real variables.
Let $\phi_n(\tau_1, \ldots ,\tau_n)$ on $\RR^n_+$ be a family of symmetric functions which we may also view as functions $\phi (e_1, \ldots , e_n)$ of the corresponding symmetric polynomials. Let $A$, $B$ and $AB$ be systems to which we associate variables $(x_1, \ldots , x_m) \in \RR^m_+$, $(x_1, \ldots , x_n)\in \RR^n_+$ and $(x_{1,1},  \ldots , x_{m,n})\in \RR^{mn}_+$ respectively. We will write $\phi_{mn}(AB)$ for 
$\phi_{mn}(x_{1,1},  \ldots , x_{m,n})$ and $\phi_m(A)$ (resp. $\phi_n(B)$) for $\phi_m$ (resp. $\phi_n$) evaluated on the {\em marginal} variable values $x_i=\sum_j x_{i,j}$ (resp. $x_j= \sum_i x_{i,j}$). For these sets of variables let the corresponding symmetric polynomials (all constructed from the $x_{i,j}$'s) be denoted by $e^{AB}_k$, $e^A_k$ and $e_k^B$, so that the symmetric functions $\phi_{mn}(AB)$, $\phi_m(A)$ and $\phi_n(B)$ may alternatively be viewed as functions of these symmetric polynomials.

\begin{lemma}\label{lemma3}
  Suppose $\phi_n$ is a family of symmetric functions as above, satisfying the following two properties for all $n$ (when taken as functions of the symmetric polynomial variables):\\
(Extendability): $ \phi_{n+1}(e_1, e_2, \ldots, e_n, 0)=\phi_n(e_1, e_2, \ldots,
  e_n)$;\\
(Monotonicity): $\partial \phi_n/\partial e_k \ge 0$, $2 \le k \le n$.\\
Then $\phi_m(A) \le \phi_{mn}(AB)$.
\end{lemma}
\noindent{\bf Proof}\, 
First note that
$
\phi_k(e_1,\ldots ,e_k) \le \phi_n(e_1, \ldots ,e_k,e_{k+1},\ldots,e_n),
$
for any $k <n$. This follows from extendability which gives
$\phi_n(e_1, ...,e_k,0,...,0)=\phi_k(e_1,....e_k)$, and then monotonicity as we increase the last $n-k$ coordinates from $(0,...,0)$
to $(e_{k+1},\ldots,e_n)$.

Next observe that the symmetric polynomial variables satisfy $e^A_1=e^{AB}_1$, and
$e^A_k \le e^{AB}_k$, for $k>1$, since every product of $x_{i,j}$ variables of the
joint system that appears in $e^A_k$ also appears in $e^{AB}_k$. Thus
\[
\phi_m(A)=\phi_m(e^A_1, \ldots ,e^A_m) \le \phi(e^{AB}_1, \ldots ,e^{AB}_m) \le \phi(e^{AB}_1, \ldots ,e^{AB}_m, e^{AB}_{m+1},\ldots e^{AB}_{mn})=\phi(AB). \, \, \,  \Box 
\] 
\begin{property}\label{propab}
\begin{align*} 
H(A) &\le H(AB),\\ 
Q(A) &\le Q(AB),\\
(-1)^{m-1} \frac{\partial^m H(A)}{\partial e_{i_1}\ldots \partial e_{i_m}} &\le (-1)^{m-1} \frac{\partial^m H(AB)}{\partial e_{i_1}\ldots \partial e_{i_m}} \mbox{  for  } m \ge 1,\\
(-1)^{m-1} \frac{\partial^m Q(A)}{\partial e_{i_1}\ldots \partial e_{i_m}} &\le (-1)^{m-1} \frac{\partial^m Q(AB)}{\partial e_{i_1}\ldots \partial e_{i_m}}  \mbox{  for  } m \ge 1.\\
\end{align*}
\hspace{14cm} $\Box$
\end{property}
This follows from Lemma \ref{lemma3} and Property \ref{prop4}. $H$ and $Q$
and their derivatives are extendable because $e_{k+1}= \ldots e_n=0$
implies that $n-k$ of the $x_i$ are zero, and both $H$ and $Q$ are
clearly extendable as functions of the $x_i$. The first inequality in Property \ref{propab} is just the well known non-negativity of conditional Shannon information and the second inequality was proved also in \cite {ddbj} by different means.

The next result is curious in that it starts from a property on $(e_1,
\ldots ,e_d)$-space and deduces one on $(x_1, \ldots , x_d)$-space.
\begin{lemma}\label{schur}
Let $f(x_1, \ldots ,x_d)$ be a symmetric function that satisfies $\partial f/\partial
e_q \ge 0$ for $q \ge 2$ (when represented as a function of the $e_k$'s). Then $f$ is Schur concave in the $x_j$ variables.
\end{lemma}
\noindent{\bf Proof}
Write $f_j=\partial f/\partial x_j$.  Viewing eq. (\ref{polypeq}) as implicitly defining $x_i=x_i(e_1, \ldots , e_d)$ we get (cf eqs. (11) and (12) in \cite{mj04}) $\partial x_j/\partial e_k=(-1)^{k+1} \sum_{j=1}^d x_j^{d-k}/\prod_{i\neq j} (x_j-x_i)$.
Then the chain rule gives
\[
\frac{\partial f}{\partial e_k}=(-1)^{k+1} \sum_{j=1}^d \frac{x_j^{d-k}f_j}{\prod_{i\neq j} (x_j-x_i)}.
\]
Now let $\hat e_i$ denote the elementary symmetric functions of the $d-2$ variables $\{x_3, \ldots, x_d\}$, with $\hat e_0=1$. Then
\begin{align*}
\sum_{q=2}^d \hat e_{q-2} \frac{\partial f}{\partial e_q} &=-\sum_{j=1}^d f_j \frac{\left[x_j^{d-2}-\hat e_1 x_j^{d-3}+ \ldots + (-1)^{d-2} \hat e_{d-2}\right]}{\prod_{i\neq j} (x_j-x_i)}\\
&=-\sum_{j=1}^d f_j \frac{(x_j-x_3) \ldots (x_j-x_d)}{\prod_{i\neq j} (x_j-x_i)}
=-\frac{(f_1-f_2)}{(x_1-x_2)}.
\end{align*}
However, $\partial f/\partial e_q \ge 0$ for $q \ge 2$ implies
$\sum_{q=2}^d \hat e_{q-2} \partial f/\partial e_q \ge 0$, so
$(f_1-f_2)/(x_1-x_2) \le 0$, which is equivalent to Schur concavity in the $x_j$ variables. \, \, \, $\Box$

\begin{property}
  $H(x_1, \ldots , x_d)$ and $Q(x_1, \ldots , x_d)$ are Schur concave.\,\, $\Box$
\end{property}
Schur concavity of $Q$ was also shown by different means in \cite {ddbj}.

Finally we make a connection with a result proved by Hellmund and Uhlmann \cite{HU}. They proved an upper bound on the von Neumann entropy of a state that specializes, for diagonal matrices, to give
\begin{align}
\label{HUbound} H(e_1,e_2, \ldots,e_d) \le -e_1\log e_1+\log d \sqrt{\frac{2d e_2}{d-1}}.
\end{align}
We can obtain similar bounds, depending only on $e_1$ and $e_2$, for
both $H$ and $Q$.
\begin{lemma}\label{hardlemma}
Let $e_k$ be the elementary symmetric functions for the set $\{x_1, \ldots, x_d\}$ and $f_k$ be those for the set $\{a, \ldots, a,b\}$, where $a$ is repeated $n-1$ times. Choose $a$ and $b$ so that $f_1=e_1$ and $f_2=e_2$, i.e. so that
\begin{align}
\label{e1ab}(d-1)a+b&=e_1,\\
\label{e2ab} {d-1 \choose 2}a^2+(d-1)ab&=e_2.
\end{align}
Choose the root 
\begin{align}\label{root}
a=\frac{(d-1)e_1-\sqrt{(d-1)^2e_1^2-2e_2d(d-1)}}{d(d-1)}.
\end{align}
Then $f_k \ge e_k$ for $1 \le k \le d$.
\hspace{14cm} $\Box$ 
\end{lemma}
\noindent {\bf Proof}\, 
We proceed by induction. Let us say that $d$ variables are a {\em canonical set} if they  consist of $d-1$ $a$'s and one $b$, with  $a \le b$. Suppose we have established that, given the constraint of any particular values for $e_1$ and $e_2$, $e_k$ is maximised only by a canonical set, for all $d$ when $k<K$, and for $d<D$ when $k=K$. Suppose $e_K$ for $d=D$ is maximised by some $x_1, \ldots ,x_D$ that is not a canonical set. Then, one can remove one of the  $x$'s, which we can assume to be $x_1$, with the remaining $x$'s remaining a non-canonical set. Then, for all $k \ge 1$ (with $e_0=1$)
\begin{align}
\label{inductive} e_k(x_1, \ldots x_D)=e_k(x_2,\ldots,x_D)+x_1e_{k-1}(x_2, \ldots x_D),
\end{align}
and since the inductive hypothesis holds for $k=K$ and $d=D-1$, the assignment $x_2, \ldots, x_D$ does not maximise either $e_K$ or  $e_{K-1}$, subject to  $e_1$ and  $e_2$ being fixed at the values $e_1(x_2, \ldots, x_D)$ and  $e_2(x_2, \ldots, x_D)$, respectively. Thus we can find $y_2, \ldots ,y_D$ such that $e_K(y_2, \ldots ,y_D)>e_K(x_2, \ldots ,x_D)$ and  $e_{K-1}(y_2, \ldots ,y_D)>e_{K-1}(x_2, \ldots ,x_D)$, respectively, while $e_1$ and $e_2$ (on these $D-1$ elements) are fixed. Thus
\begin{align*}
e_2(x_1,y_2, \ldots ,y_D)&=e_2(y_2, \ldots, y_D)+x_1e_1(y_2, \ldots, y_D)\\
&=e_2(x_2, \ldots , x_D)+x_1e_1(x_2, \ldots , x_D)\\
&=e_2(x_1, \ldots , x_D),
\end{align*}
and similarly $e_1(x_1,y_2, \ldots ,y_D)=e_1(x_1, \ldots , x_D)$.
Thus we have found a set of $D$ variables such that $e_1$ and $e_2$ are fixed and $e_K$ is larger than  $e_K(x_1, \ldots , x_D)$, which contradicts the assumption that  $e_K$ is maximised by a non-canonical set.

We need to check the initial steps of the induction. For $e_3$ with $d=3$, maximising $\sum_{i<j<k}x_ix_jx_k+\lambda(\sum_i x_i-e_1)+\mu (\sum_{i<j} x_ix_j-e_2)$, with Lagrange multipliers $\lambda$, $\mu$, gives a quadratic in the $x$'s, so we can assume there are only two possible values for the $x$'s, which we call $a$ and $b$, with $a<b$. Thus the $x$'s must be of the form $a,a,b$ or $a',b',b'$, and a simple calculation shows that only the former maximises $e_3$; this amounts to choosing the solution of eqs. (\ref{e1ab}) and (\ref{e2ab}) given by eq. (\ref{root}). For larger $d$, we suppose as before that $e_3$ is maximised by a non-canonical set $x_1, \ldots x_d$. Then eq. (\ref{inductive}) becomes
\[
e_3(x_1, \ldots x_d)=e_3(x_2, \ldots x_d)+x_1e_2(x_2, \ldots x_d),
\]
and we proceed as before, except that we find an assignment $y_2, \ldots , y_d$ that increases the $e_3$ term on the righthand side but keeps the $e_2$ term fixed. For $e_k$ with $d=k$, eq. (\ref{inductive}) takes the simpler form  $e_k(x_1, \ldots x_k)=x_1e_k(x_2,\ldots,x_k)$, and the argument proceeds as before.\,\, $\Box$

Using $\partial H/\partial e_k > 0$ and $\partial Q/\partial e_k > 0$,  Lemma \ref{hardlemma} tells us that $H$ and $Q$ can only increase when the $x_k$ are replaced by  the set $\{a, \ldots, a,b\}$. The basic definitions of $H$ and $Q$, eqs. (\ref{heq}) and (\ref{qdeq}), then give the following bounds:
\begin{property}\label{HQbounds}
\begin{align}
\label{ourHbound}H(e_1,e_2, \ldots, e_d) &\le -(d-1)a\log a-b\log b,\\
Q(e_1,e_2, \ldots, e_d) &\le \frac{d \ ! \ a^2 \log a}{2(a-b)}-\frac{b^d \log b}{(b-1)^{d-1}},
\end{align}
where $a$ and $b$, depending only on $e_1$ and $e_2$,  are given by eqs. (\ref{root}) and (\ref{e1ab}). 
\,\,$\Box$
\end{property}
These are the tightest possible bounds that depend only on $e_1$ and $e_2$ since they are
attained by a particular corresponding assignment of $x_j$'s. In particular, inequality
(\ref{ourHbound}) is tighter than (\ref{HUbound}).


\begin{thebibliography}{~~} \label{refs}

\bibitem{jrw} R.~Jozsa, D.~Robb and W.~K.~Wootters, ``Lower bound for accessible information in quantum mechanics'',
{\em Phys.~Rev.~A} {\bf 49} p668-677 (1994).

\bibitem{wkw0} W. Wootters, ``Random quantum states'' {\em Found. Phys.} {\bf 20} p1365-1378 (1990).

\bibitem{hol} A. S. Holevo, {\em Probl. Inf. Transm.(USSR)} {\bf 9} p177 (1973).

\bibitem{mj04} G. Mitchison and R. Jozsa, ``Towards a geometrical interpretation of quantum-information compression'' {\em Phys. Rev. A} {\bf 69} p032304 (2004)

\bibitem{fannes} M. Fannes, ``Monotonicity of von Neumann entropy  expressed as a function of Renyi entropies'', arXiv:1310.5941 (2013).

\bibitem{HU} M. Hellmund and A. Uhlmann,``An entropy inequality'' {\em Quantum Information and Computation} {\bf 9} p622-627 (2009).

\bibitem{bernsteinpaper} S. N. Bernstein, ``Sur les fonctions absolument monotones'' {\em Acta Mathematica} {\bf 52} p1-66 (1928).

\bibitem{Bernstein-functions} R.~L.~Schilling, R.~Song, Z.~Voncracek, ``Bernstein functions, theory and applications'', De Gruyter Studies in Mathematics, 37 (2012).

\bibitem{feller} W.~Feller, ``An introduction to probability theory
  and its applications. Volume II'', John Wiley and Sons, New York
  (1971).

\bibitem{koenraad} K. Audenaert, ``Trace inequalities for completely monotone functions and Bernstein functions'',     {\em Linear Algebra Appl.} {\bf 437(2)} p601-611 (2012).

\bibitem{bhatia} R. Bhatia, ``Matrix analysis'', Graduate Texts in Mathematics 169, Springer Verlag (1996).

\bibitem{ddbj} N. Datta, T. Dorlas, R. Jozsa and F. Benatti, ``Properties of subentropy'' (2013) arXiv:1310.1312. {\em J. Math. Phys.} (to appear).

\bibitem{Agler} J. Agler, J. E. McCarthy, N. J. Young, ``Operator monotone functions and L\"owner functions of several variables.'' arXiv:1009.3921 [math.FA]. Ann. Math. (2) {\bf 176}, 1783-1826 (2012). 

\bibitem{Hansen} F. Hansen, ``Operator monotone functions of several variables.'' arXiv:math/0205147 [math.OA]. Math. Ineq. Appl. {\bf 6}, 1-17 (2003).





\end{thebibliography}
\end{document}